\documentclass[runningheads,a4paper]{llncs}

\usepackage{amssymb}
\usepackage{graphicx}

\usepackage{url}
\urldef{\mailsa}\path{ mathcci@yahoo.com}
\urldef{\mailsb}\path {dvh@vnu.edu.vn}
\urldef{\mailsc}\path|erika.siebert-cole, peter.strasser, lncs}@springer.com|

\begin{document}

\mainmatter  

\title{SDMS-based Disk Encryption Method}%

\titlerunning{SDMS-based Disk Encryption Method}

%
%
\author{Dokjun An, Myongchol Ri, Changil Choe, Sunam Han, and Yongmin Kim }
\authorrunning{Dokjun An, Myongchol Ri, Changil Choe, Sunam Han, and Yongmin Kim}

\institute{Faculty of Mathematics, Kim Il Sung University, D.P.R.K\\
\mailsa
}

%
%

\toctitle{Lecture Notes in Computer Science}
\tocauthor{Authors' Instructions}
\maketitle

\begin{abstract}
We propose a disk encryption method, called secure disk mixed system (SDMS) in this paper, for data protection of disk storages such as USB flash memory, USB hard disk and CD/DVD. It is aimed to solve temporal and spatial limitation problems of existing disk encryption methods and to control security performance flexibly according to the security requirement of system. SDMS stores data by encrypting with different encryption key per sector and updates sector encryption keys each time data is written. Security performance of SDMS is analyzed at the end of the paper.
\end{abstract}

\section{Introduction}
Massive use of mobile storage media raises data security problem seriously. Data security of mobile storage media can be realized in different ways, e.g., file unit encryption, file system level encryption, and full disk encryption [1, 3, 4, 5, 6]. Reliability of security system should not be based on system mechanism or complexity of system analysis, and it should guarantee safety even if system mechanism or encryption algorithm is opened to third party. We analyze security problems of existing disk encryption methods in the next section and describe our disk encryption method based on SDMS in the following section. In the last section, we analyze security performance of our system.

\section{Previous Works }

Several different methods for disk encryption, such as LoopAES, EFS, TrueCrypt, NCryptfs, were suggested [4, 5, 9, 10, 11]. In these methods, encryption of disk block is expressed as follows.
\[\begin{array}{llll}
C=OP(BE(OP(P,DEK,i), DEK), DEK, i)
\end{array}\]
                      
Here, BE is block encryption function (AES, 3DES, etc), OP is operation function (CBC, LRW, XTS, etc), DEK is disk encryption key, C is ciphertext, P is plaintext and i is block index. As above expression shows, these disk encryption systems encrypt plaintext using symmetric-key algorithm through certain operation and apply this operation again to encrypted result.
[7] and [8] explain BitLocker disk encryption method and its security strength, which is offered in Windows Vista. Here, plaintext is XORed with sector key, passed two diffusers in succession, and finally encrypted using AES of CBC mode. These disk encryption methods have some weakness in terms of time passage and space expansion. We call it temporal limitation and spatial limitation, respectively, in this paper. 

\begin{itemize}
\item $\emph{Temporal limitation}$

If third party succeed to detect encryption key of a certain sector, data which is stored later in this sector can be decoded.

\item $\emph{Spatial limitation}$ 

If third party succeed to detect encryption key of a certain block, whole data of disk is in danger of being decoded.
\end{itemize}

GBDE based encryption method which was implemented in FreeBSD overcame these temporal and spatial limitations to a certain extent [6]. In this method, data is stored in sector by being encrypted with different key each time data is writen, because it generates random data newly and encrypt plaintext using it. Therefore, it is impossible to decrypt data stored newly even though third party succeeds to detect key by attacking a sector. And each key encrypting plaintext sectors are different each other when it writes data on disk, because GBDE encrypts plaintext using randomly generated key. Thus, it is impossible to decrypt data of other block even if third party detects encryption key by attacking a data sector. Although GBDE based disk encryption method overcomes temporal and spatial limitations of previous disk encryption methods considerably, it still has some security problems to be solved. 

At first, key-key for a given sector is fixed because it is decided depending on the sector address. That is, when it was writen new data on sector, sector key is encrypted by same key-key. Then, if attacker detect 128bit sector key by attacking AES/CBC/128 encrypted plaintext data and detect key-key subsequently by attacking AES/CBC/256 encrypted sector key, it is possible to decrypt newly stored data on this sector. We think this is temporal limitation of GBDE based disk encryption.

Next, it is easy to get keychain used to encrypt plaintext data, if the correlation between random data generated consecutively by PRNG is revealed, because it directly uses random data generated by PRNG as key for plaintext. However, strictly speaking, PRNG generates data deterministically based on the initial value. If attacker succeeds to get sector key by attacking key sector and subsequently succeeds to know inner state of PRNG, he can detect following sector keys easily. This allows possibility of decrypting consecutive ciphers by attacking one sector. Of course, it is possible to make difficult to predict future data from past data using cryptographically secure PRNG, but it causes another security problem that safety of system depends on the safety of PRNG too much. We think this is spatial limitation of GBDE based disk encryption. In the next section, we present a new disk encryption method based on the SDMS.

\section{SDMS-Based Disk Encryption}

In section 2, we discussed temporal and spatial limitations of previous works for disk encryption and concluded that GBDE still has security problem to be solved, while it is a good disk encryption method. In this section, we propose SDMS (secure disk mixed system) aimed to solve temporal and spatial limitation of existing disk encryption methods and to control security performance flexibly according to the security requirement of system. 

\subsection{SDMS}

SDMS is a method to encrypt each sector by generating sector key using randomly generated SEED and disk encryption key DEK. In our method, encryption key of each sector is different each other and it is changed whenever encryption is done.

\subsubsection{Data Structure}

SDMS manages data area of media by dividing into SDMS blocks. Each block consists of SDMS\_BLOCK\_DA area storing encrypted data and SDMS\_BLOCK\_SA area storing random numbers which are used to generate encryptioin key for the encryption of SDMS\_BLOCK\_DA area.
 
SDMS\_BLOCK\_SA area consists of SDMS\_UNIT\_SEEDs which are SEED data for each sector. Fig. 1 shows data structure of SDMS. Size of each area is determined depeding on the size of random data SDMS\_UNIT\_SEED needed to generate sector key. 

\[\begin{array}{llll}
SIZE(SDMS\_BLOCK\_DA) =\\

$  $ $  $ $  $ $  $ $  $ $  $ $  $ $  $ $  $ $  $ $  $ $  $   SIZE(SDMS\_BLOCK\_SA) / SIZE(SDMS\_UNIT\_SEED) \times 512
\end{array}\]

If we store SDMS\_BLOCK\_SA in one sector, the number of sector of SDMS\_
BLOCK\_DA is equal to 512 / SIZE(SDMS\_UNIT\_SEED). For example, if we set SIZE(SDMS\_UNIT\_SEED) = 8 byte (128 bit), then the number of sectors of SDMS\_BLOCK\_DA is equal to $512 \times 8 / 128 = 64$. That is, 64 plaintext sectors constitute a SDMS block (SDMS\_BLOCK) and one SEED sector (SDMS\_BLOCK\_SA) is in this block.

\begin{figure}
\centering
\includegraphics[height=5.3cm]{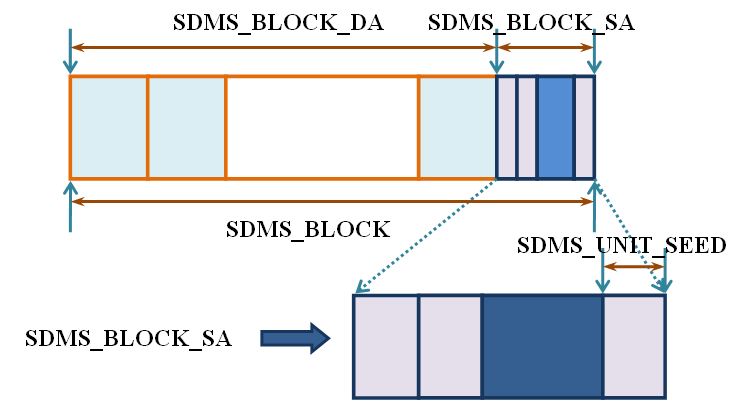}
\caption{Data structure of SDMS}
\label{fig2}
\end{figure}

There is no constraint that SDMS\_BLOCK\_SA must be one sector in SDMS block. System designer can adjust this setting freely according to the security requirement and this setting will change processing of blocks of SDMS.

\subsubsection{Data Encryption}

Encryption mode of SDMS is expressed as follows.
\[\begin{array}{llll}
C = EA (P, RTEK)
\end{array}\]

Here, EA is encryption function, RTEK is encryption key, C is cipher, and P is plaintext. RTEK is determined on the fly in time of real-time encryption (or decryption) as follows.

\[\begin{array}{llll}
RTEK = DK\_Func(DEK, SEED, i)
\end{array}\]

Here, DEK is disk encryption key, SEED is random data in SDMS\_UNIT\_SEED area, i is sector index and DK\_Func is key derivation function. DK\_Func is a function to derive real-time encryption key from disk encryption key, SEED and sector index. It is a one-way function where the length of output is constantly the same as the length of RTEK. Encryption and decryption algorithms of SDMS are as follows.

\begin{itemize}
\item $\emph{Encryption}$

\begin{itemize}
\item Writing request for i-th sector
\item Random generation of SEED
\item Calculation of RTEK
\item Encryption of plaintext with RTEK
\item Writing cipher on SDMS\_BLOCK\_DA
\item Writing SEED on SDMS\_BLOCK\_SA
\end{itemize}
\item $\emph{Decryption}$ 

\begin{itemize}
\item Reading request for i-th sector
\item Getting SEED from SDMS\_BLOCK\_SA
\item Reading cipher from SDMS\_BLOCK\_SA
\item Calculation of RTEK
\item Decryption of cipher with RTEK
\end{itemize}
\end{itemize}

\begin{figure}
\centering
\includegraphics[height=5.3cm]{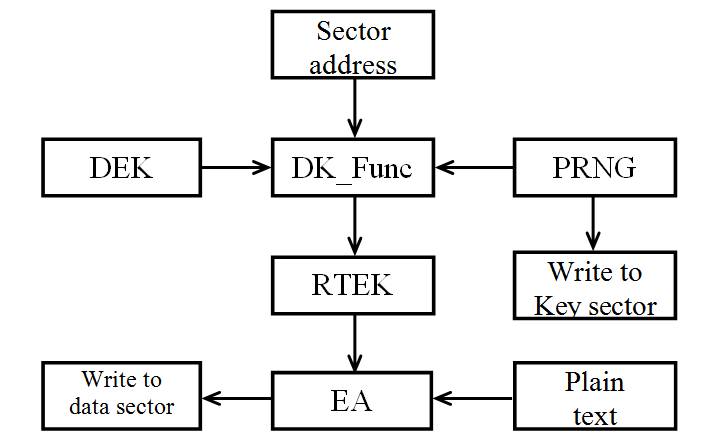}
\caption{Data encryption in SDMS}
\label{fig2}
\end{figure}

\subsection{DEK Management} 

DEK is generated when disk is initialized and is used to encrypt whole data of disk. If DEK is revealed, attacker can decrypt whole data of disk. DEK can be stored in the same disk with plaintext or in physically separated memory device such as USB memory or file server of high security level. No matter it is stored in data disk or physically isolated memory device, DEK must be encrypted based on the user authentication information. User can be authenticated through PKCS\#5 based password authentication or PKCS\#11 based smart card authentication or SSL based network authentication, but what is important is to receive key safely for the decryption of DEK [12, 13]. In this paper, we only consider disk data encryption which uses DEK, on the premise that DEK is managed safely, though DEK is very important for the reliabile management of data.

\subsection{Sector Key Derivation Function DK\_Func}

DK\_Func is a function to get sector encryption key RTEK in real time for the encryption of sector.
\[\begin{array}{llll}
RTEK = DK\_Func(DEK, SEED, i)
\end{array}\]

As we can see here, it outputs sector key for sector encryption from DEK, SEED corresponding to the sector, and the sector index. Input data space must be larger enough than output data space in the design of DK\_Func. For instance, if DEK is 2048 bit, sector index is 32bit and SEED size is 128bit, then input data space is $2^{2048+32+128}$. Therefore, in case of using AES/XTS for sector encryption, 512 bit key is needed, and hence output space is $2^{512}$ which is smaller enough than input space. DK\_Func must be implemented so as to satisfy one-wayness and collision resistance as possible. So it is desirable to construct DK\_Func using cryptographically safe hash function or hash chain.

\subsection{Security Feature of SDMS}

Firstly, SDMS solved temporal limitation problem of data encryption. SDMS generates encryption key by generating random SEED newly each time when writing request occurs. This makes it impossible to decrypt sector data written later, although attacker succeeds to break RTEK of the sector.

Secondly, SDMS solved spatial limitation problem of data encryption to a certain extent. In SDMS, even though attacker succeeds to attack EA encrypted certain sector, the only thing he knows is RTEK of that sector. He can’t read the contents of other sectors unless he knows DEK through the attack on EA or DK\_Func. In this way, SDMS overcomes remarkably security weakness that whole data of disk can be revealed by succeeding to attack a particular sector. 

Thirdly, SDMS can control security performance flexibly according to the sequrity requirements. SDMS generates encryption key using disk encryption key DEK and random SEED. DEK can be set big enough according to the security requirement. For instance, if we set DEK as bigger than 1024 bit, then search space will be increased more than $2^{1024}$ when attacker attacks DKFunc to know DEK. The size of SEED (SIZE(SDMS\_UNIT\_SEED) ) which is used to generate sector key for sector encryption can also be set big enough according to the security requirement of system and there is no restriction that SDMS\_UNIT\_SEED must be arranged to each sector. Configuring system to generate sector key by arranging one SDMS\_UNIT\_SEED to several sectors, we can coordinate balance between security performance and operation cost reasonably. 

\section{Analysis Result}

\subsection{Cost of Brute Force Search}

Attacker must attack encrypted sector data unless he knows user authentication information or decrypted DEK by evil code. If sector keys that encrypted sector data have no statistical characteristics and there is no information or algorithm helpful to estimate sector key, brute force search will be appropriate method. In case using AES/XTS/256 for sector encryption, the amount of computation will be equal to $W_{AES/XTS/256}\times 2^{512}$. Here, AES/XTS encryption is expressed as follows [16].

\[\begin{array}{llll}
C_i = E_{k1} ( P_{i} \  XOR\  ( E_{k2} (n) \times a^i ))\ XOR\ ( E_{k2}(n) \times a^i )
\end{array}\]

Here, $\times$ is multiplication operator in modulo $GF(2)=x^{128}+x^7+x^2+1$, K1 is key of symmetric key encryption algorithm (E), K2 is secondary key, i is block index in encryption unit (sector), n is address of encryption unit (sector) and α is base of GF (Galois Field). 

Attacker also can decrypt whole data of disk if he knows DEK. Therefore, attacker may try to attack DEK and then to decrypt sector data using it. To attack DEK, he must attack DK\_Func which derives sector key. He can find necessary items in $2^{SIZE(DEK)}$ space because he knows SEED and sector index which are the inputs of DK\_Func. For instance, if SIZE(DEK)=2048, attacker must search $2^{2048}$ space. In case AES/XTS/256 is selected as EA, 256bit key for XTS operation and 256bit key for final AES block encryption are needed, and thus totally needed key is 512bit. Therefore, attacker can find candidates outputting same RTEK if he calculate DK\_Func for about $2^{512}$ DEK candidates with computation $W_{DK\_Func} \times 2^{512}$. There exists about $2^{2048-512}$ candidates in this case.

Now consider the case decoding an other sector using this candidate. Assuming sector index is 32bit, the possibility of successful decryption of next sector is very small. 

\[\begin{array}{llll}
1/2^{2048-512+32+SIZE(SDMS\_UNIT\_SEED)}= \\
$ $ $ $ $ $ $ $ $ $ $ $ $ $ $ $ $ $ $ $ $ $ $ $ $ $ $ $ $ $ $ $ $ $ $ $ $ $ $ $ 2^{-(2048-512+32+SIZE(SDMS\_UNIT\_SEED))}
\end{array}\]

That is because we use sector index and SEED when we calculate RTEK for other sector. To get correct DEK from $2^{2048-512}$ DEK candidates in $2^{2048}$ space is very difficult, though it would be possible to get sector key by attacking particular sector.

\subsection{PRNG and Security Performance}

Security problem in case using the result of PRNG as key directly for sector encryption was considered in section 2. SDMS uses the output of PRNG as input of DK\_Func for getting sector key and stores output of PRNG on disk without changing. To attack disk data encrypted using SDMS needs not attack PRNG. Random data generated by PRNG

\begin{itemize}
\item solves temporal limitation of disk encryption by changing sector key each time it writes data,
\item solves spatial limitation of disk encryption by setting sector key of each sector differently, 
\item makes it more difficult to attack DK\_Func.
\end{itemize}

That is, performance and quality of PRNG in SDMS have no big relevance with security performance of system.

\end{document}